\documentstyle [12pt]{article}
\makeatletter \oddsidemargin  0in \evensidemargin 0in
\textwidth16cm \RequirePackage[dvips]{graphicx} \textheight 20.5cm
\newcommand{\singlespacing}{\let\CS=\@currsize\renewcommand{\baselinestretch}{1.5}\tiny\CS}
\makeatletter \oddsidemargin  -.1in \evensidemargin -.1in
\newcommand{\doublespacing}{\let\CS=\@currsize\renewcommand{\baselinestretch}{1.35}\tiny\CS}

\def\@citex[#1]#2{\if@filesw\immediate\write\@auxout{\string\citation{#2}}\fi
  \def\@citea{}\@cite{\@for\@citeb:=#2\do
    {\@citea\def\@citea{,\linebreak[0]\hskip0pt plus .2em}%
      \@ifundefined{b@\@citeb}%
    {{\bf ?}\@warning{Citation `\@citeb' on page \thepage\space undefined}}%
      \hbox{\csname b@\@citeb\endcsname}}}{#1}}

\newtheorem{rule-def}[theorem]{Rule}
\begin{document}
\title{\bf Partial Swapping, Unitarity and No-signalling}
\author{I.Chakrabarty\thanks{Corresponding author:
E-Mail-indranilc@indiainfo.com }\\ Department of
Mathematics\\Heritage Institute of Technology,Kolkata-107,West
Bengal,India}
\date{}
\maketitle{}
\begin{abstract}It is a well known fact that an quantum  state
$|\psi(\theta,\phi)\rangle$ is represented by a point on the Bloch
sphere, characterized by two parameters $\theta$ and $\phi$. In a
recent work we already proved that it is impossible to partially
swap these quantum parameters. Here in this work we will show that
this impossibility theorem is consistent with principles like
unitarity of quantum mechanics and no signalling principle.
\end{abstract}
\section{ Introduction }In quantum information theory
understanding the limits of fidelity of different operations has
become an important area of research. Noticing these kind of
operations which are feasible in classical world but have a much
restricted domain in quantum information theory started with the
famous 'no-cloning' theorem [1]. The theorem states that one
cannot make a perfect replica of a single quantum state. Later it
was also shown by Pati and Braunstein that we cannot delete either
of the two quantum states when we are provided with two identical
quantum states at our input port [2]. In spite of these two famous
'no-cloning' [1] and 'no-deletion' [2] theorem there are many
other 'no-go' theorems like 'no-self replication' [3] ,
'no-partial erasure' [4], 'no-splitting' [5] .\\
Recently in ref [6], we introduce a new no-go theorem, which we
refer as 'no partial swapping' of quantum information. Since we
know that the information content in a qubit is dependent on the
angles azimuthal and phase angles $\theta$ and $\phi$, then the
partial swapping of quantum parameters $\theta$ and $\phi$ is
given by,
\begin{eqnarray}
|A(\theta_1,\phi_1)\rangle|\bar{A}(\bar{\theta_1},\bar{\phi_1})\rangle\longrightarrow
|A(\theta_1,\bar{\phi_1})\rangle|\bar{A}(\bar{\theta_1},\phi_1)\rangle\\
|A(\theta_1,\phi_1)\rangle|\bar{A}(\bar{\theta_1},\bar{\phi_1})\rangle\longrightarrow
|A(\bar{\theta_1},\phi_1)\rangle|\bar{A}(\theta_1,\bar{\phi_1})\rangle
\end{eqnarray}
However in ref [6] we showed that this operation is impossible in
the quantum domain from the linear structure of quantum theory.\\
In this work we once again claim this impossibility from two
different principles namely [i] unitarity of quantum mechanics
[ii] no signalling principle. The organization of the work is as
follows: In the first section we will prove this impossibility
from the unitarity of quantum mechanics. In the second section we
will do the same from the principle of no signalling. Then the
conclusion follows.\\
\section{Partial Swapping: Unitarity of Quantum Mechanics}
Let us consider a set S consisting of two non orthogonal states
$S=\{|A(\theta_1,\phi_1)\rangle,|B(\theta_2,\phi_2)\rangle\}$ Let
us assume that hypothetically it is possible to partially swap
the  parameters of these two states
$|A(\theta_1,\phi_1)\rangle,|B(\theta_2,\phi_2)\rangle$.\\
First of all we will assume that at least in principle swapping of
phase angles of two quantum states are possible, keeping the
azimuthal angles fixed. Therefore the transformation describing
such an action is given by,
\begin{eqnarray}
|A(\theta_1,\phi_1)\rangle|\bar{A}(\bar{\theta_1},\bar{\phi_1})\rangle\longrightarrow|A(\theta_1,\bar{\phi_1})\rangle|\bar{A}(\bar{\theta_1},\phi_1)\rangle\nonumber\\
|A(\theta_2,\phi_2)\rangle|\bar{A}(\bar{\theta_2},\bar{\phi_2})\rangle\longrightarrow|A(\theta_2,\bar{\phi_2})\rangle|\bar{A}(\bar{\theta_2},\phi_2)\rangle
\end{eqnarray}
To preserve the unitarity of the above transformation, will
preserve the inner product.
\begin{eqnarray}
\langle
A(\theta_1,\phi_1)|A(\theta_2,\phi_2)\rangle\langle\bar{A}(\bar{\theta_1},\bar{\phi_1})
|\bar{A}(\bar{\theta_2},\bar{\phi_2})\rangle=\langle
A(\theta_1,\bar{\phi_1}) |A(\theta_2,\bar{\phi_2})\rangle\langle
\bar{A}(\bar{\theta_1},\phi_1)
|\bar{A}(\bar{\theta_2},\phi_2)\rangle
\end{eqnarray}
The above equality will not hold for all values of
$(\theta,\phi)$. The equality will hold if $i)
\tan\frac{\theta_1}{2}\tan\frac{\theta_2}{2}=\tan\frac{\bar{\theta_1}}{2}\tan\frac{\bar{\theta_2}}{2}$
or $ii) (\phi_2-\phi_1)=(\bar{\phi_2}-\bar{\phi_1})\pm 2k\pi$,
where $k$ is an integer. These two conditions characterize  the
set of states on the Bloch sphere for which the partial swapping
of the phase angles are possible. However in general this is not
true for all possible values of $\theta_i$, $\phi_i$ where $(i=1,2)$.\\
Let us now assume that partial swapping of azimuthal angles are
possible, without altering the phase angles of the quantum states.
\begin{eqnarray}
|A(\theta_1,\phi_1)\rangle|\bar{A}(\bar{\theta_1},\bar{\phi_1})\rangle\longrightarrow|A(\bar{\theta_1},\phi_1)\rangle|\bar{A}(\theta_1,\bar{\phi_1})\rangle\nonumber\\
|A(\theta_2,\phi_2)\rangle|\bar{A}(\bar{\theta_2},\bar{\phi_2})\rangle\longrightarrow|A(\bar{\theta_2},\phi_2)\rangle|\bar{A}(\theta_2,\bar{\phi_2})\rangle
\end{eqnarray}
Now once again, in order to preserve the unitarity of such a
transformation we arrive at the same conditions, (i) and (ii).
This clearly indicates the fact that there are certain class of
states on the bloch sphere for which partial swapping of phase
angles and azimuthal angles are possible. However in this context
we cannot say that this is true for all such values of phase and
azimuthal on the bloch sphere. Therefore it is evident that the
unitarity of quantum mechanics, doesn't allow partial swapping of
quantum parameters for all such pairs of non orthogonal states on
the bloch sphere.
\section{Partial Swapping: Principle of No signalling}
Suppose we have two identical singlet states $|\chi\rangle$ shared
by two distant parties Alice and Bob. Since the singlet states are
invariant under local unitary operations, it remains same in all
basis. The states are given by
\begin{eqnarray}
|\chi\rangle|\chi\rangle &=&
\frac{1}{2}(|\psi_1(\theta_1,\phi_1)\rangle|\bar{\psi_1}(\bar{\theta_1},\bar{\phi_1})\rangle-|\bar{\psi_1}(\bar{\theta_1},\bar{\phi_1})\rangle|\psi_1(\theta_1,\phi_1)\rangle){}\nonumber\\&&(|\psi_1(\theta_1,\phi_1)\rangle|\bar{\psi_1}(\bar{\theta_1},\bar{\phi_1})\rangle-|\bar{\psi_1}(\bar{\theta_1},\bar{\phi_1})\rangle|\psi_1(\theta_1,\phi_1)\rangle){}\nonumber\\&&
=\frac{1}{2}(|\psi_2(\theta_2,\phi_2)\rangle|\bar{\psi_2}(\bar{\theta_2},\bar{\phi_2})\rangle-|\bar{\psi_2}(\bar{\theta_2},\bar{\phi_2})\rangle|\psi_2(\theta_2,\phi_2)\rangle){}\nonumber\\&&(|\psi_2(\theta_2,\phi_2)\rangle|\bar{\psi_2}(\bar{\theta_2},\bar{\phi_2})\rangle-|\bar{\psi_2}(\bar{\theta_2},\bar{\phi_2})\rangle|\psi_2(\theta_2,\phi_2)\rangle)
\end{eqnarray}
where $\{|\psi_1\rangle, |\bar{\psi_1}\rangle \}$ and $\{
|\psi_2\rangle, |\bar{\psi_2}\rangle \}$ are two sets of mutually
orthogonal spin states (qubit basis). Alice possesses the first
particle while Bob possesses the second particle. Alice can choose
to measure the spin in any one of the qubit basis namely
$\{|\psi_1\rangle, |\bar{\psi_1}\rangle \}$, $\{ |\psi_2\rangle,
|\bar{\psi_2}\rangle \}$.\\

The theorem of no signalling tells us that the measurement outcome
of one of the two parties are invariant under local unitary
transformation done by other party on his or her qubit.The density
matrix is invariant under local unitary operation by the other
party. Hence the first party cannot distinguish two
mixtures due to the unitary operation done at remote place.\\
At this point one may ask if Alice(Bob) partially swap the quantum
parameters of her(his) particle and if Bob(Alice) measure his(her)
particle in either of the two basis then is there any possibility
that Alice(Bob) know the basis in which Bob(Alice) measures
his(her) qubit or in other words, is there any way by which
Alice(Bob) using a perfect partial swapping machine can
distinguish the statistical mixture in her(his) subsystem
resulting from the measurement done by Bob(Alice). If Alice(Bob)
can do this then signalling will take place, which is impossible.
Note that whatever measurement Bob(Alice) does, Alice(Bob) does
not learn the results and her(his) description will remain as that
of a completely random mixture , i.e.,
$\rho_{A(B)}=\frac{I}{2}\otimes \frac{I}{2}$. In other words we
can say that the local operations performed on his(her) subspace
has no effect on Alice's(Bob's)
description of her(his) states.\\
Let us consider a situation where Alice is in possession of a
hypothetical machine which can partially swap quantum parameters
$\theta$ and $\phi$.\\
 Let us first of all consider the case where
with the help of the machine we can partially swap the phase
angles keeping the azimuthal angles of the states fixed. The
action of such a machine is given by,
\begin{eqnarray}
|\psi_i(\theta_i,\phi_i)\rangle|\bar{\psi_i}(\bar{\theta_i},\bar{\phi_i})\rangle\longrightarrow|\psi_i(\theta_i,\bar{\phi_i})\rangle|\bar{\psi_i}(\bar{\theta_i},\phi_i)\rangle\nonumber\\
\end{eqnarray}
where $(i=1,2)$. Now if after the action of such a transformation
on Alice's qubit,the entangled state initially shared between
these two parties takes the form,
\begin{eqnarray}
|\chi\rangle_{PS}|\chi\rangle_{PS}&=&\frac{1}{2}[(|\psi_1(\theta_1,\phi_1)\psi_1(\theta_1,\phi_1)\rangle)_A(|\bar{\psi_1}(\bar{\theta_1},\bar{\phi_1})\bar{\psi_1}(\bar{\theta_1},\bar{\phi_1})\rangle)_B
+{}\nonumber\\&&(|\bar{\psi_1}(\bar{\theta_1},\bar{\phi_1})\bar{\psi_1}(\bar{\theta_1},\bar{\phi_1})\rangle)_A(|\psi_1(\theta_1,\phi_1)\psi_1(\theta_1,\phi_1)\rangle)_B
-{}\nonumber\\&&(|\psi_1(\theta_1,\bar{\phi_1})\bar{\psi_1}(\bar{\theta_1},\phi_1)\rangle)_A(|\bar{\psi_1}(\bar{\theta_1},\bar{\phi_1})\psi_1(\theta_1,\phi_1)\rangle)_B
-{}\nonumber\\&&(|\bar{\psi_1}(\bar{\theta_1},\phi_1)\psi_1(\theta_1,\bar{\phi_1})\rangle)_A(|\psi_1(\theta_1,\phi_1)\bar{\psi_1}(\bar{\theta_1},\bar{\phi_1})\rangle)_B]{}\nonumber\\&&
=\frac{1}{2}[(|\psi_2(\theta_2,\phi_2)\psi_2(\theta_2,\phi_2)\rangle)_A(|\bar{\psi_2}(\bar{\theta_2},\bar{\phi_2})\bar{\psi_2}(\bar{\theta_2},\bar{\phi_2})\rangle)_B
+{}\nonumber\\&&(|\bar{\psi_2}(\bar{\theta_2},\bar{\phi_2})\bar{\psi_2}(\bar{\theta_2},\bar{\phi_2})\rangle)_A(|\psi_2(\theta_2,\phi_2)\psi_2(\theta_2,\phi_2)\rangle)_B
-{}\nonumber\\&&(|\psi_2(\theta_2,\bar{\phi_2})\bar{\psi_2}(\bar{\theta_2},\phi_2)\rangle)_A(|\bar{\psi_2}(\bar{\theta_2},\bar{\phi_2})\psi_2(\theta_2,\phi_2)\rangle)_B
-{}\nonumber\\&&(|\bar{\psi_2}(\bar{\theta_2},\phi_2)\psi_2(\theta_2,\bar{\phi_2})\rangle)_A(|\psi_2(\theta_2,\phi_2)\bar{\psi_2}(\bar{\theta_2},\bar{\phi_2})\rangle)_B]
\end{eqnarray}
where A, B denotes the particles in Alice's and Bob's possession
respectively. \\
Now, if Bob does his measurement on
$\{|\psi_1\rangle, |\bar{\psi_1}\rangle \}$ qubit basis, then the
reduced density matrix describing Alice's subsystem is given by,
\begin{eqnarray}
\rho_A &=&
\frac{1}{4}[|\psi_1(\theta_1,\phi_1)\psi_1(\theta_1,\phi_1)\rangle\langle
 \psi_1(\theta_1,\phi_1)\psi_1(\theta_1,\phi_1)|+{}\nonumber\\&&|\bar{\psi_1}(\bar{\theta_1},\bar{\phi_1})\bar{\psi_1}(\bar{\theta_1},\bar{\phi_1})\rangle\langle
 \bar{\psi_1}(\bar{\theta_1},\bar{\phi_1})\bar{\psi_1}(\bar{\theta_1},\bar{\phi_1})|+{}\nonumber\\&&
 |\psi_1(\theta_1,\bar{\phi_1})\bar{\psi_1}(\bar{\theta_1},\phi_1)\rangle\langle
 \psi_1(\theta_1,\bar{\phi_1})\bar{\psi_1}(\bar{\theta_1},\phi_1)|+{}\nonumber\\&&
|\bar{\psi_1}(\bar{\theta_1},\phi_1)\psi_1(\theta_1,\bar{\phi_1})\rangle\langle
\bar{\psi_1}(\bar{\theta_1},\phi_1)\psi_1(\theta_1,\bar{\phi_1})|
]
\end{eqnarray}
Interestingly if Bob does his measurement in $\{|\psi_2\rangle,
|\bar{\psi_2}\rangle \}$ qubit basis, then the density matrix
representing Alice's subsystem is given by,
\begin{eqnarray}
\rho_A &=&
\frac{1}{4}[|\psi_2(\theta_2,\phi_2)\psi_2(\theta_2,\phi_2)\rangle\langle
 \psi_2(\theta_2,\phi_2)\psi_2(\theta_2,\phi_2)|+{}\nonumber\\&&|\bar{\psi_2}(\bar{\theta_2},\bar{\phi_2})\bar{\psi_2}(\bar{\theta_2},\bar{\phi_2})\rangle\langle
 \bar{\psi_2}(\bar{\theta_2},\bar{\phi_2})\bar{\psi_2}(\bar{\theta_2},\bar{\phi_2})|+{}\nonumber\\&&
 |\psi_2(\theta_2,\bar{\phi_2})\bar{\psi_2}(\bar{\theta_2},\phi_2)\rangle\langle
 \psi_2(\theta_2,\bar{\phi_2})\bar{\psi_2}(\bar{\theta_2},\phi_2)|+{}\nonumber\\&&
|\bar{\psi_2}(\bar{\theta_2},\phi_2)\psi_2(\theta_2,\bar{\phi_2})\rangle\langle
\bar{\psi_2}(\bar{\theta_2},\phi_2)\psi_2(\theta_2,\bar{\phi_2})|
]
\end{eqnarray}
It is clearly evident that the equations (9) and (10) are not
identical in any respect and henceforth we will conclude that
Alice can distinguish the basis in which Bob has performed the
measurement, This is impossible in principle as this will violate
the causality. Hence we arrive at a contradiction with the
assumption that the partial swapping of phase angle is possible.\\
Next we see that whether the partial swapping of azimuthal angles
is consistent with the principle of no signalling or not.If we
assume that the partial swapping of phase angles is possible
keeping the azimuthal angles fixed, then its action is given by,
\begin{eqnarray}
|\psi_i(\theta_i,\phi_i)\rangle|\bar{\psi_i}(\bar{\theta_i},\bar{\phi_i})\rangle\longrightarrow|\psi_i(\bar{\theta_i},\phi_i)\rangle|\bar{\psi_i}(\theta_i,\phi_i)\rangle
\end{eqnarray}
Let us assume that this hypothetical machine is in possession of
Alice, and she applies the transformation (11) on her particles as
a result of which the entangled state (6) takes the form,
\begin{eqnarray}
|\chi\rangle_{PS}|\chi\rangle_{PS}&=&\frac{1}{2}[(|\psi_1(\theta_1,\phi_1)\psi_1(\theta_1,\phi_1)\rangle)_A(|\bar{\psi_1}(\bar{\theta_1},\bar{\phi_1})\bar{\psi_1}(\bar{\theta_1},\bar{\phi_1})\rangle)_B
+{}\nonumber\\&&(|\bar{\psi_1}(\bar{\theta_1},\bar{\phi_1})\bar{\psi_1}(\bar{\theta_1},\bar{\phi_1})\rangle)_A(|\psi_1(\theta_1,\phi_1)\psi_1(\theta_1,\phi_1)\rangle)_B
-{}\nonumber\\&&(|\psi_1(\bar{\theta_1},\phi_1)\bar{\psi_1}(\theta_1,\phi_1)\rangle)_A(|\bar{\psi_1}(\bar{\theta_1},\bar{\phi_1})\psi_1(\theta_1,\phi_1)\rangle)_B
-{}\nonumber\\&&(|\bar{\psi_1}(\theta_1,\phi_1)\psi_1(\bar{\theta_1},\bar{\phi_1})\rangle)_A(|\psi_1(\theta_1,\phi_1)\bar{\psi_1}(\bar{\theta_1},\bar{\phi_1})\rangle)_B]{}\nonumber\\&&
=\frac{1}{2}[(|\psi_2(\theta_2,\phi_2)\psi_2(\theta_2,\phi_2)\rangle)_A(|\bar{\psi_2}(\bar{\theta_2},\bar{\phi_2})\bar{\psi_2}(\bar{\theta_2},\bar{\phi_2})\rangle)_B
+{}\nonumber\\&&(|\bar{\psi_2}(\bar{\theta_2},\bar{\phi_2})\bar{\psi_2}(\bar{\theta_2},\bar{\phi_2})\rangle)_A(|\psi_2(\theta_2,\phi_2)\psi_2(\theta_2,\phi_2)\rangle)_B
-{}\nonumber\\&&(|\psi_2(\bar{\theta_2},\bar{\phi_2})\bar{\psi_2}(\theta_2,\phi_2)\rangle)_A(|\bar{\psi_2}(\bar{\theta_2},\bar{\phi_2})\psi_2(\theta_2,\phi_2)\rangle)_B
-{}\nonumber\\&&(|\bar{\psi_2}(\theta_2,\phi_2)\psi_2(\bar{\theta_2},\bar{\phi_2})\rangle)_A(|\psi_2(\theta_2,\phi_2)\bar{\psi_2}(\bar{\theta_2},\bar{\phi_2})\rangle)_B]
\end{eqnarray}
If Bob does his measurement in any one of the two basis
$\{|\psi_1\rangle, |\bar{\psi_1}\rangle \}$ and $\{|\psi_2\rangle,
|\bar{\psi_2}\rangle \}$, then the respective density matrix
representing Alice's subsystem is given as,
\begin{eqnarray}
\rho_A &=&
\frac{1}{4}[|\psi_1(\theta_1,\phi_1)\psi_1(\theta_1,\phi_1)\rangle\langle
 \psi_1(\theta_1,\phi_1)\psi_1(\theta_1,\phi_1)|+{}\nonumber\\&&|\bar{\psi_1}(\bar{\theta_1},\bar{\phi_1})\bar{\psi_1}(\bar{\theta_1},\bar{\phi_1})\rangle\langle
 \bar{\psi_1}(\bar{\theta_1},\bar{\phi_1})\bar{\psi_1}(\bar{\theta_1},\bar{\phi_1})|+{}\nonumber\\&&
 |\psi_1(\bar{\theta_1},\phi_1)\bar{\psi_1}(\theta_1,\bar{\phi_1})\rangle\langle
 \psi_1(\bar{\theta_1},\phi_1)\bar{\psi_1}(\theta_1,\bar{\phi_1})|+{}\nonumber\\&&
|\bar{\psi_1}(\theta_1,\bar{\phi_1})\psi_1(\bar{\theta_1},\phi_1)\rangle\langle
\bar{\psi_1}(\theta_1,\bar{\phi_1})\psi_1(\bar{\theta_1},\phi_1)|
]{}\nonumber\\&&
=\frac{1}{4}[|\psi_2(\theta_2,\phi_2)\psi_2(\theta_2,\phi_2)\rangle\langle
 \psi_2(\theta_2,\phi_2)\psi_2(\theta_2,\phi_2)|+{}\nonumber\\&&|\bar{\psi_2}(\bar{\theta_2},\bar{\phi_2})\bar{\psi_2}(\bar{\theta_2},\bar{\phi_2})\rangle\langle
 \bar{\psi_2}(\bar{\theta_2},\bar{\phi_2})\bar{\psi_2}(\bar{\theta_2},\bar{\phi_2})|+{}\nonumber\\&&
 |\psi_2(\bar{\theta_2},\phi_2)\bar{\psi_2}(\theta_2,\bar{\phi_2})\rangle\langle
 \psi_2(\bar{\theta_2},\phi_2)\bar{\psi_2}(\theta_2,\bar{\phi_2})|+{}\nonumber\\&&
|\bar{\psi_2}(\theta_2,\bar{\phi_2})\psi_2(\bar{\theta_2},\phi_2)\rangle\langle
\bar{\psi_2}(\theta_2,\bar{\phi_2})\psi_2(\bar{\theta_2},\phi_2)|
]
\end{eqnarray}
 Alice can easily distinguish two
statistical mixtures and as a consequence of which she can easily
understand in which basis Bob has performed his measurement. This
is not possible in principle as this will violate causality.
Hence we conclude that the partial swapping of azimuthal angles
is not possible.
\section{Acknowledgement}
I.C acknowledges Almighty and Prof. C.G.Chakrabarti for being the
source of inspiration in carrying out research.
\section{References}
$[1]$ W.K.Wootters and W.H.Zurek,Nature \textbf{299},802(1982).\\
$[2]$ A.K.Pati and S.L.Braunstein, Nature \textbf{404},164(2000).\\
$[3]$ A.K.Pati and S.L.Braunstein, e-print quant-ph/0007121.\\
$[4]$ A.K.Pati and Barry C.Sanders, Physics Letters A 359, 31-36 (2006).\\
$[5]$ Duanlu Zhou, Bei Zeng, and L. You, Physics Letters A 352, 41 (2006).\\
$[6]$  Indranil Chakrabarty, e-print quant-ph/0612123.(Accepted in
IJQI).

\end{document}